\documentclass[12pt]{article}
\pdfoutput=1
\usepackage{amsfonts}
\usepackage{url}
\usepackage{hyperref}
\usepackage{simpler-wick}
\usepackage{bbold}
\usepackage{slashed}
\usepackage{graphicx}
\usepackage{pgfplots}
\usepackage{color}
\usepackage{mathrsfs}
\usepackage{fancybox}
\usepackage{lipsum}
\usepackage{eurosym}
\usepackage{tcolorbox}
\usepackage{lmodern}
\usepackage{tikz}
\usepackage{tikz,pgfplots}
\usepackage{pgfplots}
\usepackage{amsmath,amsthm,amssymb}
\pgfplotsset{compat=1.10}
\usepgfplotslibrary{fillbetween}
\usepackage{feynmp}
\usepackage{slashed}
\usepackage{graphicx}
\usepackage{epstopdf}
\usepackage{subfigure}
\usepackage{color}
\usetikzlibrary{decorations.pathmorphing}
\tikzset{snake it/.style={decorate, decoration=snake}}
\usetikzlibrary{shapes}
\usepackage{empheq}
\usepackage{mathrsfs}
\usepackage{lmodern}

\def\({\left (}
\def\){\right )}
\def\[{\left [}
\def\]{\right ]}

\numberwithin{equation}{section}

\def\lsim{\mathrel{\rlap{\lower4pt\hbox{\hskip1pt$\sim$}}
     \raise1pt\hbox{$<$}}}         
\def\gsim{\mathrel{\rlap{\lower4pt\hbox{\hskip1pt$\sim$}}
     \raise1pt\hbox{$>$}}}         

\begin{document}
\newcommand{\nnn}{n}
\def\smallTFD{{\rm TFD}}

\addtolength{\abovedisplayskip}{.8mm}
\addtolength{\belowdisplayskip}{.8mm}
\addtolength{\parskip}{.8mm}
\begin{titlepage}

\setcounter{page}{1} \baselineskip=15.5pt \thispagestyle{empty}

\vfil

${}$
\vspace{1cm}

\begin{center}

\def\thefootnote{\fnsymbol{footnote}}
\begin{center}
{\Large \bf A Conversation on ER = EPR}
\end{center}

~\\[1cm]
{Erik Verlinde\footnote{\href{mailto:E.P.Verlinde@uva.nl}{\protect\path{E.P.Verlinde@uva.nl}}} and Herman Verlinde\footnote{\href{mailto:verlinde@princeton.edu}{\protect\path{verlinde@princeton.edu}}}}
\\[0.3cm]

{\normalsize { \sl Institute of Physics,  
University of Amsterdam,\\ 
Science Park 904, 1198 XH, Amsterdam, the Netherlands}} \\[3mm]

{\normalsize { \sl Physics Department,  
Princeton University, Princeton, NJ 08544, USA}} \\[3mm]

\end{center}



{\small  \noindent 
\begin{center} 
\textbf{Abstract}\\[4mm]
\parbox{15 cm}{Ten years ago this week, the two authors had an email conversation about black holes, ER bridges, and EPR entanglement. This brief note contains a verbatim translation of these emails. While the ideas expressed in this email dialogue linking \cite{epr} and \cite{er} are more mainstream now than they were back then, there are still many  unresolved puzzles, some of which are discussed in this old correspondence. }

\end{center} }
 \vspace{0.3cm}
\vfil
\begin{flushleft}

\today
\end{flushleft}

\end{titlepage}

\subsection*{Introduction}

In the past 15 years, the relationship between EPR entanglement and the connectedness of space-time has been a  central guiding theme for research on quantum gravity and the holographic duality. Elements of these ideas were already contained in the early work on black hole radiation by Bekenstein, Hawking \cite{bh}, Unruh, Davies \cite{unruh-davies},
and others, and within the context of the AdS/CFT correspondence, in work by Maldacena \cite{eternal-ads-bh}, Ryu, Takayanagi \cite{RT}
and in papers by van Raamsdonk \cite{vanraamsdonk}. They became more prominent ten years ago as a counterpoint to the firewall paradox of Almheiri, Marolf, Polchinski and Sully~\cite{amps}.

The ER = EPR conjecture \cite{ER=EPR} is the bold statement that a large amount of entanglement between two localized regions of space-time implies the existence of a macroscopic geometric connection between two regions of space-time. As with many new ideas in science, it emerged from a period of active collective exploration,  discussions, and sharing of insights.

The following email dialogue took place ten years ago this week.  Apart from translation from the original Dutch into English, the email transcript is unedited.\footnote{The original email exchange is found \href{https://verlinde.scholar.princeton.edu/sites/g/files/toruqf3026/files/documents/Gmail-ER-EPR-bellen-merged.pdf}{here}.} 
The below thoughts about entangled black holes, Einstein-Rosen bridges, and the conceptual link between \cite{epr} and \cite{er} were shared in informal conversations with others during the subsequent weeks and months. Parallel ideas  were publicized six months later in the courageous paper \cite{ER=EPR} by Maldacena and Susskind and have become the subject of active debate since.

This brief note is written in a different format than the typical arxiv paper. We decided to post it because we believe that the email dialogue is still interesting to read today and forms an instructive illustration of how informal  communication contributes to the emergence of new scientific ideas. Several of the questions raised in the dialogue are still open points of discussion today.

\subsection*{Email correspondence with \bf Subject: EPR black holes}

\noindent
{\sf\underline{E.P.Verlinde@uva.nl  To: herman.verlinde@gmail.com
Tue, Dec 11, 2012 at 11:52 AM}}\\[-1mm]

\noindent
Hello Herman,\\
I'm at work right now, but I'm about to go home (by bike). Then I also want to be able to do some quick shopping. So for me, 7:30 probably works out. Is that OK?

\medskip

I agree with you that the solution of the paradox is only interesting if we learn something new from it. But my reason for working on this is that I believe this is indeed the case, and that in particular the need for the additional degrees of freedom and the breakdown of naive low energy QFT will become even more apparent as a result.
Then it is certainly interesting to extend these conclusions to Rindler space. In that case it seems almost inevitable that the degrees of freedom have to do with both sides of the horizon.

\medskip

Another thought I had yesterday about your idea of considering an EPR pair of maximally entangled black holes. Is such a thing possible in principle? My conclusion is that it is precisely the Einstein-Rosen bridge that should tie the two together. Because then the radiation coming from one black hole is equal to the Hawking pair of the radiation from the other black hole. I'm not sure if it makes sense, but it might be interesting to think about.

\bigskip

\noindent
{\sf\underline{herman.verlinde@gmail.com To: E.P.Verlinde@uva.nl, Tue, Dec 11, 2012 at 12:49 PM}}\\[-1mm]

\noindent
Hi Erik,\\ 
The Einstein-Rosen bridge is an interesting observation:
because strictly speaking, one can't distinguish the maximally mixed AMPS density matrix from a black hole that is completely entangled with another black hole. The properties of a density matrix should not depend on the choice of purification. 

\smallskip

So the ER bridge is the geometrification of the entanglement.

\smallskip

From this point of view, the bridge's short lifespan may reflect decoherence: the long range entanglement decoheres and gets replaced by short-distance entanglement.

\smallskip

I'll go back to the IAS in a minute: Joe is giving a second talk, this time on conformal versus scale invariance.

\smallskip

\noindent
BTW: I just completed the calculation of the "Page curve" (within our ergodic model) with my student Steven.\footnote{This calculation of Page's formula \cite{page} is described in Steven Jackson's \href{https://verlinde.scholar.princeton.edu/sites/g/files/toruqf3026/files/documents/SRJ-pre-thesis-with-added-comment.pdf}{pre-thesis}. } I'll send you a formula later.

\bigskip

\noindent
{\sf \underline{E.P.Verlinde@uva.nl  To: herman.verlinde@gmail.com,
Tue, Dec 11, 2012 at 7:44 PM}}\\[-1mm]

\noindent
We have previously established the link between EPR correlations and the ER bridge. But now I see even more clearly why this should be the case. The identification of the qubit behind the horizon and the qubit outside the black hole is exact! There is no difference. This is what is now called the $A=R_B$ scenario (although I find $H_B = R_B$ a more consistent notation). This means that the qubits that Alice finds behind the horizon are also the same as the ones that Bob (probably "later", whatever that means) will be able to detect. So it just has to be that everything beyond the horizon gets teleported out itself at some point.

\medskip

 So the funny thing is that there is no causal way to go through the ER bridge, but you can "tunnel" through it. This tunneling (a la Parikh-Wilczek${}^{[9]}$) is comparable to quantum teleportation. But even without an EPR black hole pair you actually have to make an identification on the space that identifies the region behind the horizon, as it were, with the places where the corresponding qubits are present in the external radiation. I see this as tunneling using the ``hidden" phase space, which is responsible for the decoherence of the space and time degrees of freedom.
The "assistance" that these degrees of freedom give to the teleportation protocol via the classic bits must play an important role in this.

\medskip
\bigskip

\noindent
{\sf\underline{herman.verlinde@gmail.com To: E.P.Verlinde@uva.nl, 
Tue, Dec 11, 2012 at 8:47 PM}}\\[-1mm]

\noindent
I was just trying to call - let me know if you're still up.

\medskip

About your image: you only get an ER bridge if all the entanglement is collected within a compact object (another black hole) far away from the nearby black hole. Otherwise, the ER bridge starts inside the black hole on one end, while the other end is
thinly spread out over all the radiation. The latter situation is a less clear story... But it is certainly useful to see whether the teleportation story can indeed be cast in such a geometric form.

\medskip
\bigskip

\noindent
{\sf \underline{E.P.Verlinde@uva.nl  To: herman.verlinde@gmail.com,
Wed, Dec 12, 2012 at 12:00 AM}}\\

There is a nice but also important generalization of the idea. The argument must be robust under unitary transformations of one of the two black holes. If you take this into account you can in fact handle all situations, because, as I said, time evolution but also translations count as unitary transformations.
If you perform a unitary transformation then the state is still maximally entangled, but now as
$$
\sum |i\rangle \otimes U |i\rangle
$$
So in this case you need to identify the left side of one black hole by U times the right side of the other. Note that there is also a translation in between. But you can undo this, and then they will become identical. In that case you get exactly the thermal double, and you have therefore created an eternal black hole. But this one indeed has a Hilbert space that's twice as large, as Gerard already realized.

But you can also let one of the two black holes evolve further in time. Then you get the usual AMPS situation. However, our conclusion remains unchanged. Finally, it is also logical to consider measurements and for example the quantum teleportation protocol. Then you are actually rejuvenating the black hole by means of the measurement. So this story can of course also be added.

\bigskip
\bigskip

\noindent
{\sf\underline{herman.verlinde@gmail.com To: E.P.Verlinde@uva.nl, Wed, Dec 12, 2012 at 4:19 AM}}\\

An effective way to tell the story, perhaps, is to simply imagine that there is a hypothetical universe in which all black holes have smooth horizons. We can call this the ``Twin Universe'', because it is in a maximally entangled EPR state with an identical twin. This is similar to elliptic de Sitter, but where the elliptic $Z_2$ is not divided out but indicates an EPR correlation -- and as I type this, I'm thinking that elliptic de Sitter space *should* work like this, because how else would one show that the dS horizon is smooth?! The two halves share maximum entanglement entropy, and this is how the smooth horizon is created. The world then indeed looks like Gerard's swiss cheese space-time\footnote{See figure 7 of \cite{tHooft:1984kcu}. In 't Hooft's paper, the two worlds are viewed as the two sides of a density matrix. Here they are seen as the two sides of an entangled state.}. This has in general no consequences, except for black holes and for cosmological horizons: there the entanglement is so large that there is a classical geometry associated with it.

For black holes, this means that the horizon carries $S_{BH}/\log 2$ classical bits, which may be used to reconstruct the interior operators. Or equivalently -- as you write -- you could describe this as an eternal black hole with double the number of states, but with the elliptic EPR identification imposed.

Furthermore, we can now simply evaporate black holes in this world. I then realized that Gerard's bra and ket idea is not completely wrong, because the direction of time on the other side has been reversed. In other words, the evaporation time step takes place via\footnote{The notation here follows the author's paper \cite{VV-QEC} on quantum error correction and black hole entanglement: $|i\rangle$ and $|j\rangle$ denote basis states of the internal black hole Hilbert space and $|n\rangle$ are basis states of the emitted radiation. $C^i_{jn}$ are assumed to be random matrices with the property that $\sum_{i}  C^i_{jn} C^*{}^i_{\tilde j,\tilde n}\propto \delta_{n\tilde n} \delta_{j\tilde j}$. This relation also plays a role in the Hayden-Preskill-Kitaev-Yoshida decoding protocol \cite{haydenpreskill,KitaevYoshida} and in the Gao-Jafferis-Wall quantum teleportation protocol \cite{GJW}.}
\begin{equation}
\sum_{i,jn,\tilde{j},\tilde{n}} C^i_{jn} |j\rangle|n\rangle \otimes\, C^*{}^i_{\tilde j,\tilde n}|\tilde j\rangle|\tilde n\rangle
\nonumber
\end{equation}
Now do the sum over i, and use our averaging. This gives
\begin{equation}
\sum_{j,n} | j \rangle |n\rangle \otimes | j \rangle | n \rangle
\nonumber
\end{equation}
So the maximal entanglement remains, precisely because of interference = decoherence. The Hawking particles in both worlds are indeed EPR partners. The first notation indicates that you can quantum teleport them using the classical bits "i". The quantum teleportation
protocol naively requires a lot of work: you have to measure the entire black hole state "i", and forward that information. That seems impossible. But our claim is that the ER bridge executes this protocol automatically. In other words, by setting up the story like this, we use the classical geometry of the ER bridge to argue the QT mechanism instead of the other way around.

\medskip
Finally: in our story, the perspective on the AMPS entanglement has changed considerably. An AMPS opportunist might notice that if you were to make measurements on one of the two worlds, the EPR correlation with the twin world disappears. So then our mechanism doesn't seem to work anymore. But one can then answer that the Born rule projection need not be imposed, and that the many worlds interpretation makes equivalent predictions. Or rather, since the same measurement also takes place in the twin world, you may/must perform the Born projection simultaneously. And then everything continues to work -- and as you pointed out, this is exactly how you're making EPR pairs of young black holes.

\medskip

Basically, we're almost done with the story now -- so let's just start writing it up. We still have to think about how to make this clear in pictures. The eternal black hole and Gerard's swiss cheese picture are clear. But do we have another QT protocol picture?

\bigskip
\bigskip

\noindent
{\sf\underline{herman.verlinde@gmail.com To: E.P.Verlinde@uva.nl, Wed, Dec 12, 2012 at 8:31 AM}}\\

\noindent
Here's an obvious point of worry:

Maybe depending on how causality works for the ER bridge, then of course there's the problem that any little disturbance in the distant past gets hugely blue-shifted into a firewall singularity at the future horizon of the EPR partner black hole. So are we not showing that there must be a firewall? Or is the EPR interference just such that this problem exactly cancels? In other words, is this instability precisely equal to the required mirror image of the Hawking radiation?
This does not seem very easy to argue.

\bigskip
\bigskip

\noindent
{\sf \underline{E.P.Verlinde@uva.nl  To: herman.verlinde@gmail.com, Wed, Dec 12, 2012 at 10:02 AM}}\\

\noindent
I had also thought of this: an infalling shell of matter, if boosted into the past so that it runs along the past horizon, would indeed lead to a firewall for the EPR partner.
But not entirely for the AMPS reason. There is indeed another side of the horizon where there is a copy of the EPR twin. 

\bigskip
\bigskip

\noindent
{\sf \underline{E.P.Verlinde@uva.nl  To: herman.verlinde@gmail.com, Wed, Dec 12, 2012 at 7:32 PM}}\\

\noindent
I was out tonight (competition with the chess club) but I have thought further.

   In short, I think we know how to adjust the AMPS postulates so that there is no problem with smooth horizons. You simply have to formulate more clearly what the infalling and outside observer see as common reality, and in which their experiences are complementary. The main point is that the qubit beyond the horizon may be identified with the qubit in the early radiation, and because the outside observer cannot figure out this qubit until it can be transferred to the infalling observer. One way to ``guarantee" this is to postulate that ``black holes are the fastest quantum computers". So the quickest way to extract the qubit from the early radiation is to put it in a black hole and wait for it to emit the entangled qubit. This happens at a space-like separated point in space-time. 
   
   The funny thing is that this ``twin-black hole" does indeed use the quantum error correction protocol. Namely, the transition amplitude is exactly the complex conjugate of the original C coefficient, so you find that the ancilla is equal to the bit emitted by the twin black hole. But only in this case
one applies error correction to the radiation outside the black hole that is trapped inside the twin black hole, and this QEC is reliable after the Page time.

\subsection*{Concluding Remarks}

Why was publishing these ideas not an obvious step? While there were and are many indications that entanglement and geometric connectedness are related as concepts, postulating the equivalence between the two notions is in fact a radical and unconventional statement. Unlike all observable properties of quantum systems, entanglement is not a linear quantity and not measurable in a single copy of a quantum system. So entanglement by itself can not be enough to guarantee the presence of a measurable quantity like classical smoothness of space-time. Even at the time, it was clear to the authors that additional quantum error correcting properties of the underlying quantum state and dynamics are needed to ensure that connectedness is a robust property of space-time \cite{VV-QEC,Verlinde:2012,EWR}. 

As noted in the email exchange, the eternal two-sided black hole geometry is unstable under small perturbations in the far past, which rapidly get blue-shifted into singular shockwaves along the future horizon of the EPR partner black hole \cite{DraytHooft,Kiemvv}. These shockwaves are now better understood as the geometric manifestation of the underlying quantum chaos \cite{fastscrambler}\cite{MSS} and the butterfly effect \cite{ssbutterfly}. Maintaining phase coherence between two chaotic quantum systems, so that one can perform a controlled quantum teleportation protocol \cite{GJW}, is indeed a challenging task. So rather than an obstruction to the conjectured equivalence, the instability of two-sided black hole space-time under the formation of shockwaves is an intrinsic feature that further supports the correspondence, while also illustrating the delicate nature of ER bridges and of their microscopic realization.

\section*{Acknowledgements}
\vspace{-1mm}

We thank many colleagues for inspiring discussions and insightful comments,

\addtolength{\baselineskip}{-.3mm}

\end{document}